# An ensemble deep learning technique for detecting suicidal ideation from posts in social media platforms

Shini Renjith *, Annie Abraham, Surya B. Jyothi, Lekshmi Chandran, Jincy Thomson

*Department of Computer Science and Engineering, Mar Baselios College of Engineering and Technology, Thiruvananthapuram, Kerala, India*

ABSTRACT

Suicidal ideation detection from social media is an evolving research with great challenges. Many of the people who have the tendency to suicide share their thoughts and opinions through social media platforms. As part of many researches it is observed that the publicly available posts from social media contain valuable criteria to effectively detect individuals with suicidal thoughts. The most difficult part to prevent suicide is to detect and understand the complex risk factors and warning signs that may lead to suicide. This can be achieved by identifying the sudden changes in a user's behavior automatically. Natural language processing techniques can be used to collect behavioral and textual features from social media interactions and these features can be passed to a specially designed framework to detect anomalies in human interactions that are indicators of suicidal intentions. We can achieve fast detection of suicidal ideation using deep learning and/or machine learning based classification approaches. For such a purpose, we can employ the combination of LSTM and CNN models to detect such emotions from posts of the users. In order to improve the accuracy, some approaches like using more data for training, using attention model to improve the efficiency of existing models etc. could be done. This paper proposes a LSTM-Attention-CNN combined model to analyze social media submissions to detect any underlying suicidal intentions. During evaluations, the proposed model demonstrated an accuracy of 90.3% and an F1-score of 92.6%, which is greater than the baseline models.

## 1. Introduction

The idea of suicide is considered as a susceptibility to end one's life mainly because of depression, with a high concentration with self-delusion. On the report of the World Health Organization (WHO), it is shown that every year around 800,000 people die of suicide with at least as many suicide attempts. Suicide is the second emerging cause of death among the younger generation with a suicide rate of 10.5 per 100,000 people (World Health Organization, 2018).

According to some studies (Saxena et al., 2015), (i) Health factors (e.g., mental health, chronic pain), (ii) Historical Factors (e.g., previous suicide attempts, family history), and (iii) Environmental Factors (e.g., stressful life events, harassment) are the three major risk factors of suicide. Besides, the time period preceding a suicide can hold clues to an individual's struggle. In accordance with some studies, most of the people with suicidal tendency do not attempt suicide. For example, Klonsky et al. (Klonsky and May, 2014) suspect that most of the frequently cited risk factors like frustration, stress, depression etc. related to suicide are the predictors of suicide ideation and not the progression from ideation to attempt. Nevertheless, Pompili et al. (Pompili et al., 2014) says that suicide ideator and suicide attempter are similar to "several variables assumed to be risk factors of suicidal behavior". In countries which are members of WHO, early detection of suicide ideation has been developed and implemented as a national suicide prevention strategy (World Health Organization, 2018).

Over the past few years, it is seen that social media has been a powerful "window" for the mental health and well-being of the users, mostly the youths. Mental health related forums along with social media have become a leading study area in Computational Linguistics. It gives a valuable exploration for the development of new technological improvements and approaches that can bring a difference in suicide detection and further to prevent suicidal risks (Marks, 2019). We propose to use this means to detect users who have suicidal ideation using various deep learning techniques. Apart from conventional text classification techniques, deep learning techniques have made further development in the domain of Pattern Recognition and Computer Vision. While Machine Learning has its own drawbacks, neural networks based on vector representations can bring out better outcomes on several Natural Language Processing (NLP) tasks (Yang et al., 2018). We intend to explore the capabilities of Long Short Term Memory (LSTM), Convolutional Neural Network (CNN), Attention Layer and their combined model. We hope that this immense quantity of data on people's thoughts, opinions, feelings etc. can help in early detection of suicidal posts and thus may even prevent deaths.

* Corresponding author.
*E-mail address:* shinirenjith@gmail.com (S. Renjith).

The main objectives of this work are:

1. To predict the posts that are suicidal based on Human Behavior – The model predicts about the posts based on the words or sentences written by human beings
2. To increase the accuracy of the prediction – The model predicts on each post with more accuracy since we employ an attention model which focuses on the important aspects from the obtained information along with LSTM and CNN combined model.

The major contributions of this work include:

1. Suicide ideation detection using LSTM-attention-CNN model.
2. Reads posts of specific users and determines whether the user is suicidal
3. Predictions of suicidal tendency of user is determined by analyzing posts giving importance to new posts but also taking into account old posts.

This paper is constructed as follows: Section 2 covers the related papers on suicide detection on social media. Section 3 explains the data collection process and how the dataset is modified for this project. Section 4 specifies the proposed model and Section 5 elaborates the methodology used and describes each approach adopted in the proposed model. It also mentions the additional modifications and/or steps involved. Section 6 focuses on experimental results which includes data analysis results and classification result analysis. Finally, Section 7 provides the conclusion and also mentions the directions for future work.

## 2. Literature review

In recent years, a large number of attempts were made to highlight the influencing potential of social media on suicide ideation. One such experiment is Facebook Social Media for Depression Detection in the Thai Community (Katchapakirin et al., 2018). This paper provides a tool to detect depression on Facebook, to psychologically assess and build a depression detection algorithm in Thai language by applying Natural Language Processing Techniques. Depressive signs were drawn out by using the user's behavior on Facebook which includes the user's interaction with others, number of posts, the day and time of posting and the user's privacy settings.

For the study, they selected over the internet, Facebook users, aged over 18 years and who were prepared to distribute their micro blogs for depression research. Thirty five of the selected users were divided by the Thai Mental Health Questionnaire (TMHQ) into two groups of which twenty two were classified as depressed and thirteen as non-depressed. They collected 1105 posts and extracted the attributes. Several Machine Learning (ML) algorithms like Random Forest (RF) (Donges, 2019), Deep Learning (DL) algorithms (Shrestha and Mahmood, 2019) and Support Vector Machine (SVM) (Zhang, 2012) were applied. For SVM Algorithm, the dimension of the gathered data was too small, that they were not able to divide the training set and test set as conventional validation without the loss of significant modeling. The evaluation of the model prediction performance was done using the eight-cross validation. The SVM model had an accuracy that was slightly greater than that of the majority vote which formed the baseline for evaluation.

For Random Forest Algorithm, to get a better insight, they used RapidMiner which is a Data Science software platform that gives an integrated environment for the techniques like machine learning, text mining, data preparation, deep learning, etc. The accuracy of this model was much better compared to the SVM model. For Deep Learning Algorithm, the sentiment polarity was broken down into two different groups – the complete set of posts with positive emotions and the complete set of posts with negative emotions. This prediction had confidence of about 62.64%. The most significant assistance for this prediction was from the negative sentiment. They found that people who posted without using any emoticon on negative sentimental micro blogs and also with the privacy set to 'only me' were found to be suffering from depression. On the contrary, those users who regularly forwarded others' posts, shared their memory, tagged friends, and actively posted from 6 AM to 12 AM, were possibly not suffering from depression. The accuracy of this model was slightly better compared to the other 2 models.

These experiments showed that depression can be predicted by analyzing the user's pattern on Facebook in the form of activities and messages. As Facebook has limitations to get the personal details of users, the method for getting consent became more complicated. So, the sample used in this research is comparatively little and the conclusions obtained from the research may not incorporate every important factor. Some errors might occur, since for analyzing the process, the language-related characteristics should be translated from Thai to English, because of which certain crucial sentiment polar terms might be cut out in the translation operation.

Jiahui Gao, Qijin Cheng, and Philip L. H. Yu introduced a model to detect YouTube comments showing risk for suicide (Gao et al., 2018). In this paper, YouTube comments belonging to Cantonese language (a mixture of traditional Chinese spoken in Hong Kong and English) are checked to see if they show any suicidal risk or not. Here they have made use of a single comment from a person to detect if he has any suicidal tendency rather than a series of comments belonging to a person, which is usually taken. A Deep Learning method is used for this purpose, and they have introduced focal loss in addition to resampling to solve the imbalance present in the dataset (non-risk comments were much greater). The dataset consisted of 5051 comments of which 251 were at-risk comments and 4800 were non-risk comments. They experimented in two ways, by making the dataset balanced and then training the model and by training the model with imbalanced data. For a balanced dataset, the cross-entropy loss was used and focal loss for an unbalanced dataset. In order to address the class imbalance, the loss, that is allocated to the well-classified data, is down weighted using the focal loss. It has parameters to adjust the importance of suicidal and non-suicidal inputs as well as hard and easy examples. Geometric mean was used as the performance measure.

The models were designed in two ways. In the first case, the dataset after being balanced (set A) was used to train the classifiers SVM, RF, AdaBoost (Chengsheng et al., 2017) and LSTM (Hochreiter and Schmidhuber, 1997) (in two ways, using seed filter and without using it) in which LSTM with no seed filter performed better with a g-mean of 84.3%. In the second case, the raw dataset without being balanced (Set B) was selected for training the LSTM model with seed filter and focal loss as the loss function which achieved a g-mean of 84.5%. This paper shows the increased performance of LSTM than other Machine Learning techniques with sequential data. Focal loss was also introduced which dealt with the imbalance dataset. But since the YouTube comments were in the Cantonese language, there was difficulty in analyzing it which reduced the accuracy of the model.

There is another study, Detection of Suicidal Intent in Spanish Language Social Networks using Machine Learning where its aim is to detect suicidal intention posts by Spanish speaking people (Valeriano et al., 2020). In order to implement this study, they had created a new dataset in Spanish language using twitter API obtaining a total of 100,000 posts out of which a dataset of 2068 text sentences were obtained that was annotated by humans. It was seen that 24% of the dataset was annotated as suicidal and 76% as non-suicidal. On this dataset they have performed some of the pre-processing steps like Removal of URLs, special character and numbers, Tokenization, Anonymization of tweets that contain names and Removal of stop words and have converted to text vector representation using TF-IDF, Word Embedding and Word2vec-Mean for Sentence Embeddings.

They used the Classification algorithms like Support Vector Machine (SVM) and Logistic Regression (LR) (Maalouf, 2011) and both were compared. Since there were unbalanced categories of suicidal and non-suicidal posts, they randomly took 500 non-suicidal posts from the 76% of the non-suicidal posts making it balanced. After training the model they got an accuracy of 79% of Logistic Regression and 74% for Support Vector Machine. Since the dataset contains less data for training, the accuracy is not as high as it is for English language.

Another important work in this area is on Detecting Suicidal Ideation on Reddit (Aladağ et al., 2018). In this paper; publicly available Reddit posts were used as the dataset. Posts from subreddits like Suicide Watch, Depression, Anxiety, and Shower Thoughts posted between September 2008 and October 2016 were downloaded, using Google Cloud BigQuery. Majority of the users on Subreddit are depressed and thinking about attempting suicide. Random posts from all Subreddits were selected and then manually annotated. The model is trained with random forest, ZeroR, logistic regression and SVM.

This research aims at building a model that detects long passages and blog posts that contains suicidal fantasies through text mining methods to help the administration from preventing potential suicide attempts. Results shows that it is possible to detect people thinking about suicide and provide them with appropriate support. Applying this type of suicidal related posts detection model can save thousands of lives if it is carried out properly.

An automated model for analyzing and estimating the suicidal risk of a person from that person's social media data was proposed in 'Natural Language Processing of social media as screening for suicide risk' (Coppersmith et al., 2018). The model aims to improve the existing screening methods for suicide risk found in the present health care system. The dataset was derived from two main sources, one from OurDataHelps.org users who willingly donated their digital data from wearables, social media platforms and other technologies, and also a set donated posthumously after their suicide by their loved ones and another set from users on social media who publicly discuss their past suicide attempts. A user with the gender and age (nearly) same as that of each user who attempted suicide was found to serve as a control user. The data is restricted to only contain posts from suicidal users 6 months prior to their suicide attempt and the final data contains 197,615 posts of users who attempt suicide in the next 6 months, and 197,615 posts from the corresponding controls.

This model uses pretrained GloVe embedding whose result is processed by a bi-directional LSTM (Brownlee, 2017). The output from each layer is then combined produce a single vector using skip connections into a self-attention layer (Hore and Chatterjee, 2019). A linear layer with a SoftMax function predicts the probability of the user being suicidal. 10-fold cross-validation across pairs of users were used to evaluate the classification performance. The model was optimized to detect trait-level risk for suicide, as opposed to state-level that is more related to a short period of risk. The model had limitations when we consider its dataset. The data is derived predominantly from females of the age group 18 to 24. Though results show that it works reasonably well for males belonging to a similar age group, there is not sufficient evidence to evaluate the performance from other age groups. The data mainly contains posts from users who have survived through their attempts so it may not accurately reflect the people who would die by suicide.

In the paper, Exploring and Learning Suicidal Ideation Connotations on Social Media with Deep Learning (Sawhney et al., 2018), Deep Learning models like LSTM, Vanilla RNN (Liu et al., 2016), and C-LSTM (CNN and LSTM combined model) (Zhou et al., 2015) were compared which are used to detect suicidal posts. For this purpose, the data was collected from Suicide web forums, user posts from the micro-blogging websites, Tumblr, and Reddit and also tweets were retrieved by Twitter REST API. The final dataset consists of 5213 text sentences from different tweets that were then manually annotated. On evaluation C-LSTM was found to outperform the other models. On conducting error analysis, it was found that subtle hints towards suicidal ideation were not identified by the model. There was uncertainty about the label of certain tweets for both the annotator and the model and the dataset lacked in certain suicidal phrases and needed update. The comparison of above mentioned work is briefly explained in Table 1.

## 3. Data

### 3.1. Dataset

We train our classification model with reddit dataset where users express their views and opinions via submissions. They interact through comment threads attached with every submission or post of the different users. We used the University of Maryland Reddit Suicidality Dataset built by Philip Resnik (Zirikly et al., 2019), for our research, that contains a list of non-suicidal and suicidal submissions. User's privacy is preserved by replacing personal data of the user by a unique ID. The dataset was drawn out from 2015 Full Reddit Submission Corpus, to identify (anonymous) users who might have suicidal tendencies, using posts from SuicideWatch (SW) subreddit. These reddit posts were labelled using 4 divisions which includes no risk, low risk, moderate risk, and severe risk outlined as:

- No Risk (or "None"): No possibility that the user has suicidal tendency.
- Low Risk: Some factors might be present here depicting that this user might have suicidal tendency, but chances are very low
- Moderate Risk: There are chances that the user has a tendency to attempt suicide.
- Severe Risk: There is high chance that the user will attempt to suicide.

Some of the examples of no risk, low risk, moderate risk, and severe risk is given in Table 2. The dataset contains posts from more than 11,000 users who have posted at least once on the

**Table 1**
Comparison of related works.

| Paper | Model | Evaluating Metrics |
|---|---|---|
| Facebook Social Media for Depression Detection in the Thai community (2018) (Katchapakirin et al., 2018) | Weka with SVM algorithm<br>RapidMiner with Random Forest algorithm<br>RapidMiner with Deep Learning algorithm | Accuracy – 68.57%<br>Accuracy – 84.6%<br>Accuracy – 85% |
| Detecting Comments Showing Risk for Suicide in YouTube (2018) (Gao et al., 2018) | Cross Entropy Loss with SVM – no seed filter<br>Cross Entropy Loss with SVM – seed filter<br>Cross Entropy Loss with AdaBoost – no seed filter<br>Cross Entropy Loss with AdaBoost – seed filter<br>Cross Entropy Loss with RF – no seed filter<br>Cross Entropy Loss with RF – seed filter<br>Cross Entropy Loss with LSTM-no seed filter<br>Cross Entropy Loss with LSTM-seed filter<br>Focal loss with LSTM-no seed filter<br>Focal loss with LSTM-seed filter | Geometric Mean – 78.3%<br>Geometric Mean – 78.4%<br>Geometric Mean – 79.2%<br>Geometric Mean – 78.6%<br>Geometric Mean – 74.3%<br>Geometric Mean – 69.7%<br>Geometric Mean – 84.3%<br>Geometric Mean – 82.3%<br>Geometric Mean – 81.8%<br>Geometric Mean – 84.5% |
| Detection of Suicidal Intent in Spanish Language Social Networks using Machine Learning (2020) (Valeriano et al., 2020) | SVM<br>Logistic Regression | Accuracy – 74%<br>Accuracy – 79% |
| Detecting Suicidal Ideation on Forums and Blogs: Proof-of-Concept Study (2018) (Aladağ et al., 2018) | RF<br>ZeroR<br>SVM<br>Logistic Regression | Accuracy – 89%<br>Accuracy – 50%<br>Accuracy – 92%<br>Accuracy – 92% |
| Natural Language Processing of Social Media as Screening for Suicide Risk (2018) (Coppersmith et al., 2018) | Bidirectional LSTM with Self Attention | AUC – 89% |
| Exploring and Learning Suicidal Ideation Connotations on Social Media with Deep Learning (2018) (Sawhney et al., 2018) | RNN<br>LSTM<br>C-LSTM | Accuracy – 73.7%<br>Accuracy – 78.7%<br>Accuracy – 81.2% |

**Table 2**
Sample Reddit posts of different classes.

| Label | Reddit posts |
|---|---|
| **No Risk** | ● So this little guy melted my heart when I saw him online, found out no one posted him here...<br>● ZOINKS! I just realized how much I'm starting to look like shaggy (scooby doo) help with a new haircut? |
| **Low Risk** | ● I have had a bad childhood. I wish no child to experience is this<br>● Some time ago I wanted to change my life. I wanted to be fit, I wanted to pass exams. Bit it all was only a words. Real motivation doesn't come from movies you watched or books you read. Everything is in you and I need 3 years to realize that. |
| **Moderate Risk** | ● I lied to get myself out of a mental hospital for suicidal behavior Just so I can go home and get it right this time.x-posted to r/suicidewatchEDIT:If I don't reply, it's because I'm asleep. I'm detoxing off of alcohol.<br>● Im so sorry this is happening to you. Everytimei hear something like this i wish i could be of help. But im incapable. I dont want to help anyone. I just want to die peacefully...fast. |
| **Severe Risk** | ● Looks like.. even alkoholwont help my pain now.. that fucking sucks... I will even worse when am drunk... and am afraid that i will say something bad to my friends about me... iquess its time to give up..Edit:Am rly sad about that.. even when i know that my dad was alkoholic..<br>● I tried really hard to do the deed today but in the end I couldn't |

subreddit, SuicideWatch along with few others who did not. Every post has:

a. post id: A distinct ID for the submission.
b. user id: A distinct numeric ID depicting the user who wrote the post.
c. timestamp: Time in which the post was created.
d. subreddit: The name of the subreddit or discussion forum where the submission appeared.
e. post title: Heading of the submission.
f. post body: Body of the submission.

The dataset contains two disjoint sets (Shing et al., 2018):

a. crowd: This annotation is made by crowd sources using CrowdFlower.
b. experts: This annotation is done by experts.

Each of which has 2 csv files one being all the data related to a post and other being the labels given to each user. In order to ensure more accurate labels various measures like participation on CrowdFlower was restricted to high performance annotators, long and short instructions were provided to the annotators depending on whether they were an expert or a crowd annotator, etc. were used. The short instructions identified four main risk factors, thoughts, feelings, logistics and context. The quality of the annotation across and within groups of crowd sources and experts were also done (Shing et al., 2018) which showed "cautious optimism" regarding the ability of crowd annotators to make risk assessment judgments. Some of the limitations for this dataset are they are restricted to only Reddit posts and since health records, outcomes, or even self-report questionnaires of users whose posts were taken were not available due to privacy issues, validate clinician assessments could not be used nor can provide any clinical evidence for improved validity using the assessment instructions.

We have removed the control users from the dataset for our model and combined both the crowd and experts' data.

### 3.2. Data annotation

In this dataset each user gets a label instead of each post where each user has multiple posts. So, this becomes difficult for training our model since there are very few users. At first, we assume that all the posts of a particular user get the label assigned to that user in the training sets and we create a temporary dataset. After which, pre-processing techniques are applied. Then we extract all the common words or word combinations using n-gram analysis, grouped according to labels to which we assign weights.

In order to assign weights, we attempted various methods. At first, we tried to classify the words into four labels using clustering through which we did not get the required results. It was classified more on the basis of sentiments than on the basis of suicidal predictions. Then we tried to use the Sentiment Analysis library to which we made the required changes but again no matter how much changes we made classification was purely based on sentiments. Both these methods did not give us the required results. Then we used the NLP Technique TF-IDF Vectorizer that evaluates how relevant a word is to a document in a collection of documents. But again, the result was not as we expected. It gives more importance to the words that are least occurring in the document. So, we looked into the inside of the TF-IDF library that is the base code of TF-IDF Vectorizer. We collected the most common words and then applied TF-IDF to assign weights for each word. Then using these weights, we analyzed the net weight of a particular post. After so many trials and tests we decided on threshold values that categorizes each post into their respective labels. Thus, obtaining the required dataset for our model with 760 users and a total of 69,600 posts. After further processing, it had 14,849 posts belonging to "No Risk", 13,691 posts belonging to "Low Risk", 13,462 posts belonging to "Moderate Risk" and 13,678 posts belonging to "Severe Risk" categories thus summing up to 55,680 posts. Each post had around one to three hundred sentences and around one to three thousand words.

## 4. Proposed model

To identify suicidal posts from Reddit, we employ the combination of the LSTM and CNN (Saha, 2018) model along with the attention model. The proposed model takes the resultant vector from the LSTM as an input value for the Attention layer, and the result from the attention layer as input for the convolutional layer. The Fig. 1 depicts, the proposed LSTM-Attention-CNN joined model classifies input posts into four classes representing different level of suicidal or non-suicidal tendencies.

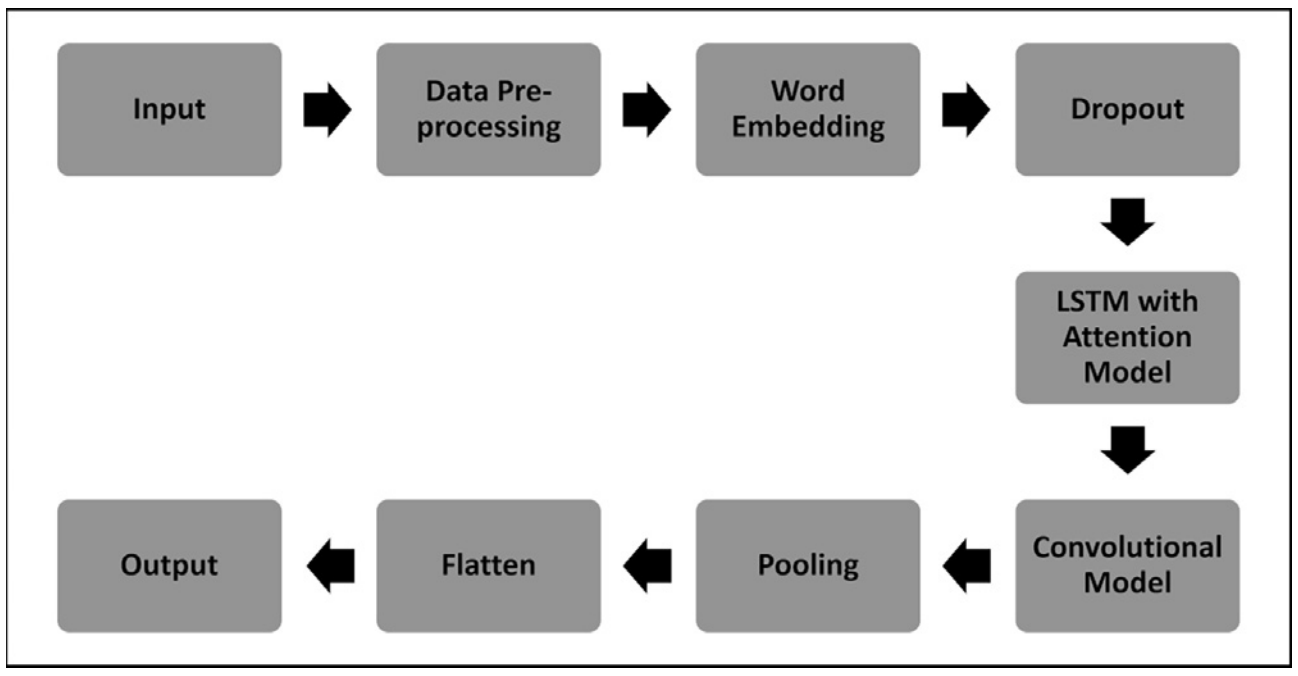

**Fig. 1.** LSTM-Attention-CNN Model for Suicidal Ideation Detection.

As shown in the figure, the first layer of the model is a word embedding layer where every token or word in each sentence is mapped to a unique index thus forming a real-valued vector of a particular length. To prevent over-fitting a dropout layer is used. LSTM layer identifies long-distance dependencies over the textual input and the convolutional layer which helps in feature extraction. The attention model highlights important information and assigns weights to each word. The pooling layer summarizes the various features present in a region in the feature map thus down sampling the feature map. Flatten layer will convert the information collected by the pooling layer into a column vector. Finally, the output layer classifies the input posts with the help of the SoftMax function.

## 5. Methodology

### 5.1. Model

This study aims to generate an implementation of a combined deep learning classifier to enhance the performance of text classification for the detection of suicidal ideation in Reddit social media. In our experiment, we use the combination of LSTM and CNN (Tadesse et al., 2019) to detect suicidal related posts. The LSTM and CNN layers can obtain information from the text but cannot concentrate on the relevant data from the information that was received. Concentrating on the relevant parts would help to increase the accuracy of the classification. Attention mechanisms can help to achieve this. It can highlight the essential data by fixing various weights. Since all the words in a sentence have varying contributions to the emotion of the context, providing various weights to different words can help improve the understanding of the emotion underlying in the whole text.

In our model, we have used the Attention layer in between the LSTM and Convolutional layer. This enables only the needed or important information selected by the LSTM layer and attention layer to pass into the convolutional layer. We have used the attention layer before the convolutional layer as it helps consider the relationship existing between different parts of the weighted input values. Thus, patterns recognized by the convolutional layer would include important patterns which might otherwise have been lost.

### 5.2. Pre-processing

Pre-processing is the process where the input textual content is filtered to eliminate noise from the raw input prior to the learning of the embedding so as to increase the effective accuracy of the proposed system. This can be achieved by using techniques like data cleansing, integration, transformation, and reduction. If data is not preprocessed, then the data would be filled with unwanted content that does not contribute to the working of the model. Noise can include punctuations, newline characters, URLs, etc. which should be removed to improve performance. Each component of the preprocessing step contributes to better output. This way the complexity of the model can be reduced as it can identify patterns quickly.

In this work, we used the Natural Language Toolkit (NLTK) (Bird et al., 2009) and other basic operations to pre-process and equip the input dataset for the usage in the training stage. First, we concatenated the post titles and post bodies. We have then removed duplicated inputs belonging to the dataset. Then, we substitute every instance of e-mail, URLs and punctuation with a single white space and also lowercase the text. Subsequently, as one component of the data filtering and conversion procedure, we apply tokenization to split the concatenated posts into a set of tokens to which normalization is applied. We removed stop words provided in the NLTK package and the newline characters, which if ignored would result in incorrect outputs. We applied lemmatization with the post tag to assure that word endings are not irregularly omitted as in stemming, which would generate senseless word items. Finally, the cleansed dataset is input into the embedding layer of the model.

### 5.3. Word embedding

Word embedding is the depiction of words in a real-valued vector format which encodes the word meaning, such that the words with similar meaning are nearer in the vector space. In the proposed LSTM-Attention-CNN joined model, word embedding layer acts as the input layer. On applying word embedding technique, each word of the vocabulary is mapped to a lower-dimensional vector space which contains real numbers (Mikolov et al., 2013). In our model, we use Word2vec (Mikolov et al., 2013), which is a more recent shallow model (two neural layers). This model is trained to recreate the context of the word. A simple example depicting the word embedding process is shown in Fig. 2.

### 5.4. Dropout layer

The purpose of this layer is to prevent overfitting and the co-adaptation of hidden units by the random dropping out of noise that may be present in the training data. The rate parameter which can have values between 0 and 1 (Srivastava et al., 2014), is given a value 0.5 where value between 0.5 and 0.8 is considered as the good value for dropout in a hidden layer. When we use the dropout operation after the embedding layer, random turning off of the activation of neurons takes place, where every neuron present in the embedding layer represents an intense depiction of a submission.

### 5.5. Long Short Term Memory

Long Short Term Memory network (LSTM), can be defined as a type of RNN, which has the capability to learn long term dependencies, which means LSTM can remember information for a longer period. The hidden layers in LSTM consist of four gates that decide which part of the context should be carried forward and how much of the past should be forgotten. Thus, it creates LSTM as a great solution for the recognition of suicidal context from social media posts. Another quality of LSTM is the prevention of explosion or the vanishing gradient that is usually found in RNN models (Brownlee, 2017).

A single layer with 150 units is applied to our LSTM layer. In each one of these cells, four gates perform four distinct calculations. An LSTM network has the input vector $[h_{t-1}, x_t]$ at time step t and the network cell state is depicted as ct. The output vectors passed through the network between consecutive time steps $t$, and $t+1$ are denoted by $h_t$. An LSTM network has three types of gates, the forget gate, input gate and output gate, that help revise and control the cell states. The gates use hyperbolic tangent and sigmoid activation functions. The LSTM executes certain pre-calculations before producing an output (Zhang et al., 2018). The Fig. 3 shows the basic LSTM network.

The equations for these gates are:

$$f_t = \sigma(W_f x_t + U_f h_{t-1} + b_f) \tag{1}$$

$$i_t = \sigma(W_i x_t + U_i h_{t-1} + b_i) \tag{2}$$

$$o_t = \sigma(W_o x_t + U_o h_{t-1} + b_o) \tag{3}$$

$$u_t = tanhtanh(W_u x_t + U_u h_t + b_u) \tag{4}$$

$$c_t = f_t^{\circ} c_{t-1} + i_t^{\circ} U_t \tag{5}$$

$$h_t = o_t(c_t) \tag{6}$$

In above mentioned equations, depicts multiplication done element wise. W, U and b, are used to depict the two weight matrices, and the bias vector respectively, and is used in the forget gate $f_t$(Eq. (1)) which is same as that for the input gate $i_t$ (Eq. (2)), $c_t$ (Eq. (5)) represents the memory cell, $u_t$ (Eq. (4)) the tanh layer, $o_t$ (Eq. (3)) depicts the output gate and $h_t$ (Eq. (6)) the hidden state. Then $h_t$ and $c_t$ values are passed for the next LSTM cell for computations. Thus, this layer gives an output sequence of values, that is $P = [p_0, p_1, p_2, \cdots, p_n]$ and then it is passed to the Attention layer.

### 5.6. Attention layer

The attention layer is used to help concentrate on different important features from the input sequence (Galassi et al., 2021). The layer takes the output from the LSTM layer and then magnifies or diminishes the value of each feature of the input based on certain activation weights, that is, we provide a higher weight to regions we wish to magnify and a lower weight to those we wish to diminish. We have used a self-attention layer with the weights, $\alpha$ being calculated using tan hyperbolic function and normalized using the SoftMax function (Eq. (7)).

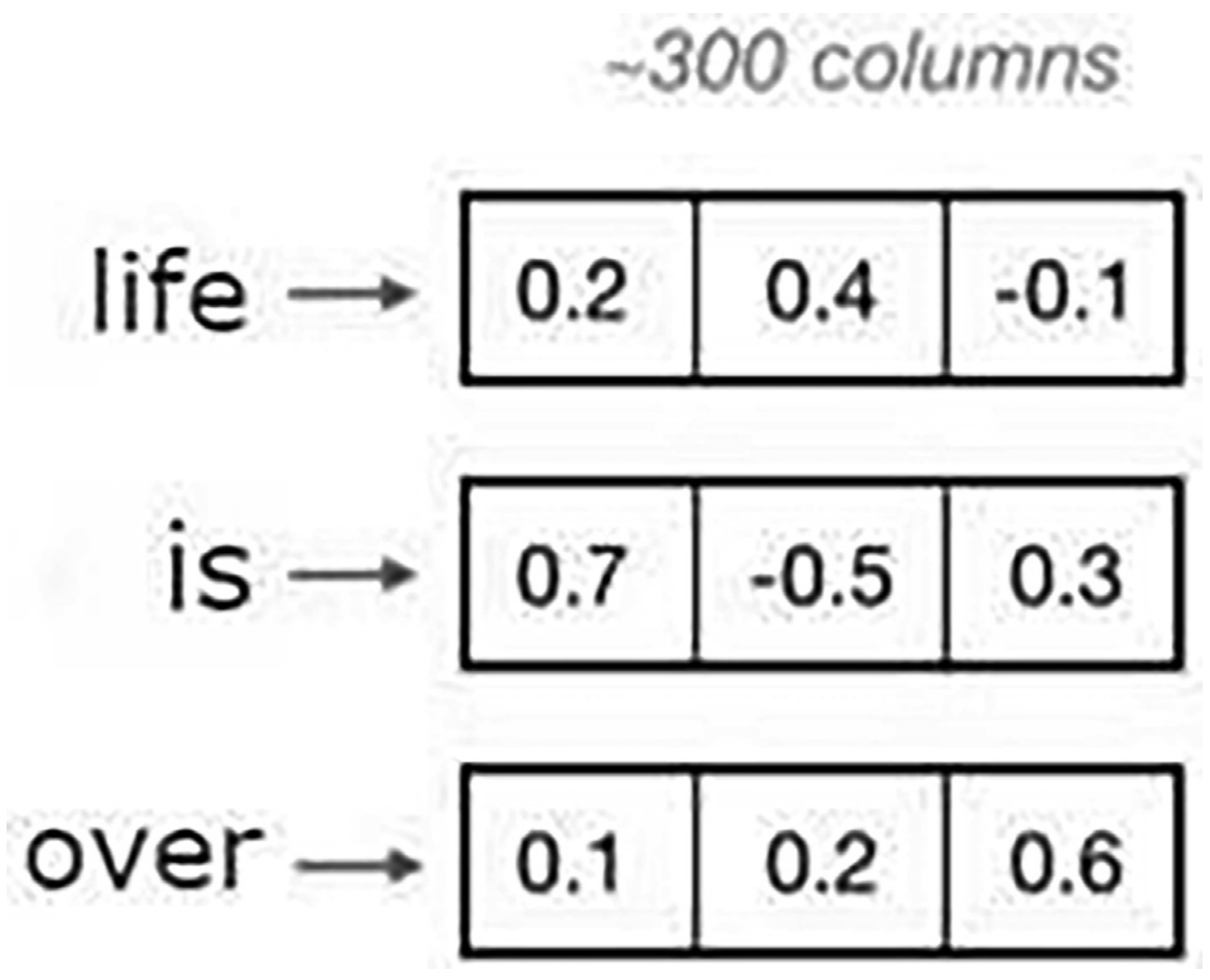

**Fig. 2.** Word embedding example.

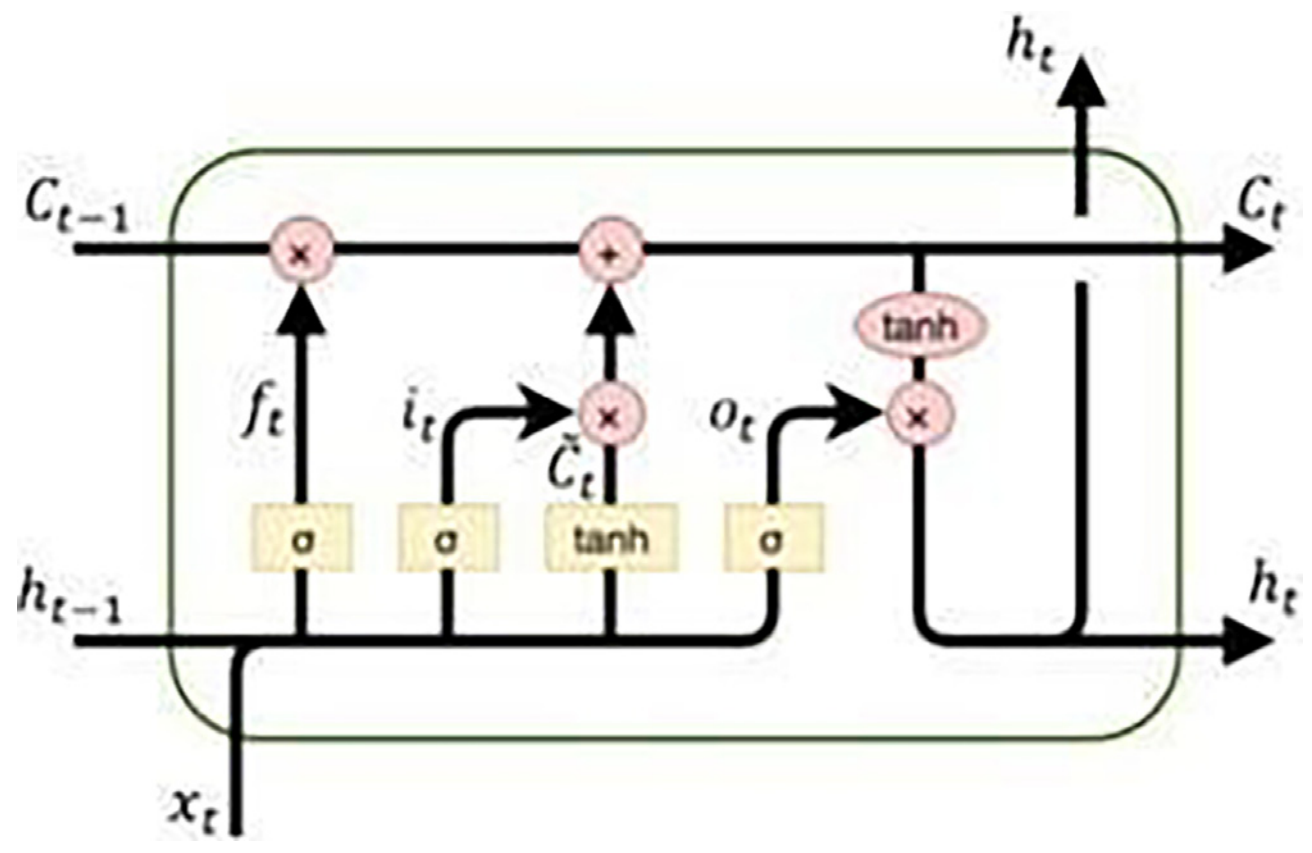

**Fig. 3.** LSTM network.

$$\alpha = softMax(tanhtanh(w \cdot P + b)) \tag{7}$$

$$output = P \times \alpha \tag{8}$$

The product of the input and the attention weight is then returned (Eq.8). Here we do not calculate the weighted sum as we take in 3D tensors as input and return the resultant 3D tensors to the next convolutional layer. This helps maintain information about the relationships between elements in the input sequence. The weighted value is then provided as input for the convolutional layer. The code snippet containing the above mentioned process of attention layer is given in Fig. 4.

### 5.7. Convolutional layer

On the input value of this layer, we apply the convolutional operation (Tadesse et al., 2019). We specify the number of filters, the kernel size, the type of padding needed and the activation function to be used. The convolution layer fires when a specific pattern is recognized. These patterns include the common n-grams found in the data, the sizes of which could be varied by changing the size of the kernels. The output generated will be similar value for different sets of similar patterns. The convolutional operation is depicted in the Fig. 5.

This layer first generates feature map with the patterns recognized based on the filters and the kernel size and then applies the activation function on the feature map to provide non-linearity. In our model we have used 3 filters with kernel size 8, with 'same' padding employed so as to maintain the dimensions of the output as same as that of the input and a RELU activation function.

### 5.8. Pooling layer

The purpose of the pooling layer is to decrease dimensionality of every rectified feature map and retain the most important data. Smaller representation of input and more manageable mass information are the characteristic features of the pooling layer. It has the ability to control over-fitting by reducing the number of parameters and the estimations present in the model (Xu et al., 2015). In order to represent only the most significant data from each feature map, we used the max pooling operation in our experiment.

### 5.9. Flatten layer

The flatten layer takes the pooled feature map as input and then flattens it using a reshape function to form a column vector. The reshape function is used for making the feature vector pulls concatenated. This forms the input value for the dense layer for the classification task (Ahmad et al., 2019).

### 5.10. Output layer

The foremost purpose of the output layer or the fully connected layer is the estimation of the possibility of a post being suicidal and non-suicidal. To prevent gradient explosion or vanishing problems, the output layer utilizes a text feature vector obtained as a result of the convolutional, pooling, and flatten functions succeeded by a substantial activation function. We may use the Hyperbolic tangent function, Sigmoid function (Norouzi et al., 2009), Rectified Linear Unit (ReLU) (Nair and Hinton, 2010), or the SoftMax function (Goodfellow et al., 2016) since the input is classified into binary. In this project, SoftMax activation is used in the output layer.

Fig. 6 depicts the code of the proposed model. Here the input given to the Embedding layer includes the dimensions of the pre trained word embedding given in input_dim and output_dim, the embedding matrix is given in the parameter weights and the length of each post of the padded dataset (here it is the length of the post with maximum length) is given through input_length.

### 5.11. Baseline

For the purpose of analyzing the proposed model we have selected a set of baseline models against which the performance

```
class Attention(Layer):

    def __init__(self, return_sequences=True):
        self.return_sequences = return_sequences
        super(Attention,self).__init__()

    def build(self, input_shape):

        self.W=self.add_weight(name="att_weight", shape=(input_shape[-1],1),
                               initializer="normal")
        self.b=self.add_weight(name="att_bias", shape=(input_shape[1],1),
                               initializer="zeros")

        super(Attention,self).build(input_shape)

    def call(self, x):

        e = K.tanh(K.dot(x,self.W)+self.b)
        a = K.softmax(e, axis=1)
        output = x*a

        if self.return_sequences:
            return output

        return K.sum(output, axis=1)
```

**Fig. 4.** Code snippet for attention layer.

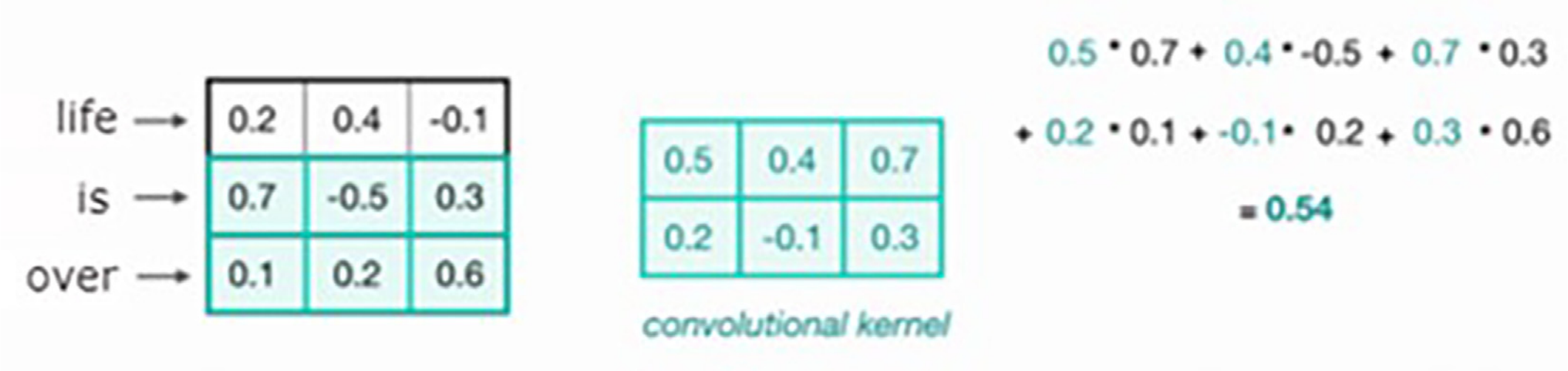

**Fig. 5.** Convolution operation on text.

```
model = Sequential()
model.add(Embedding(input_dim=word2vec.vectors.shape[0], output_dim=word2vec.vectors.shape[1],
                    weights=[word2vec.vectors], input_length=train_X1.shape[1]))
model.add(Dropout(0.5))
model.add(LSTM(units=100,return_sequences=True))
model.add(Attention(return_sequences=True))
model.add(Conv1D(3, (8,), padding='same', activation='relu'))
model.add(MaxPooling1D(2))
model.add(Flatten())
model.add(Dense(4, activation='softmax'))
```

**Fig. 6.** Code snippet for the proposed model.

of our model can be compared. The baseline models considered include some of the machine learning classifiers like:

a. TF-IDF Vectorizer – TF-IDF Vectorizer is widely used in extracting information and text mining. It measures the importance of a word in a document.
b. Support Vector Machine – Support vector machine is an algorithm that finds the separation between hyperplanes based on the classes of data. It can perform well with large feature sets as it measures the margin of separation of the data instead of matching features.
c. LSTM – LSTM is a type of recurrent neural network which is better, in terms of memory, than the traditional recurrent neural networks thus helping to maintain relevant data and discard irrelevant data.
d. CNN – Convolutional layer applies convolutional operation on the data to produce feature maps which contains information about the patterns in the data.
e. Combined LSTM-CNN model – This model has the LSTM layer to detect the long-distance dependencies along with the convolutional layer for generating feature maps.

All of these models are trained using the entire training dataset and word embedding model word2vec. All parameters for these baseline models are same (for LSTM, CNN, and LSTM-CNN) as that is used with our proposed model. Though our main focus is on the accuracy, the other evaluation metrics are also calculated.

### *5.12. Model architecture and its parameters*

In our model, we used a 300-dimensional pre-trained word2vec (Mikolov et al., 2013) model from Google News. The dataset which had 69,600 posts is split into training and test set by using 80% of the data for training (55,680 posts) and 20% for testing (13,920 posts). Most of the data was used for training so as to increase the information on which the model is trained. For compilation we provided:

(i) optimizer as Adam:

We chose Adam optimizer because the bias-correction operation in Adam helps in optimization especially towards the end when the gradients become sparse and also helps in faster convergence.

(ii) loss function as sparse categorical cross-entropy:

Sparse categorical cross entropy loss function is used when there are two or more label classes which are provided as integers.

(iii) metrics which helps estimate the performance of the model as accuracy:

Accuracy is used as an evaluation measure mainly because it is the metric which is widely used. We observed that almost all the recent literature referred by us used it and so it would help compare the results. It is also used as it is a simple metric used on balanced datasets. However, we also evaluated our model using Precision, Recall and F1-score.

ReLU helps in faster training and doesn't suffer from vanishing gradients and is found to be better for convolutional layers. Softmax activation is used in the fully connected layer since this is a multi-class classification problem. It assigns decimal probabilities to each class such that the decimal probabilities add up to 1.0. This helps in convergence.

The number of epochs 10 was selected after going through various related models. Batch size was not specified so by default the size of 32 was taken. Other parameters used in the model architecture are mentioned below in Table 3. These values including dropout rate, LSTM units, kernel size and padding were selected based on observations found in (Ahmad et al., 2019). We used Python and Anaconda framework for our experimental setup.

**Table 3**
Parameters of our model.

| Layers | Parameters | Values |
|---|---|---|
| Embedding | Embedding dimension | 300 |
| Dropout | Dropout | 0.5 |
| LSTM | Units | 100 |
| Convolutional Layer | Filters | 3 |
| | Kernel Size | 8 |
| | Padding | same |
| | Activation Function | ReLU |
| Remaining Layers | Pooling size | Max-Pooling |
| | Number of Epochs | 10 |
| | Fully Connected Layer | SoftMax |

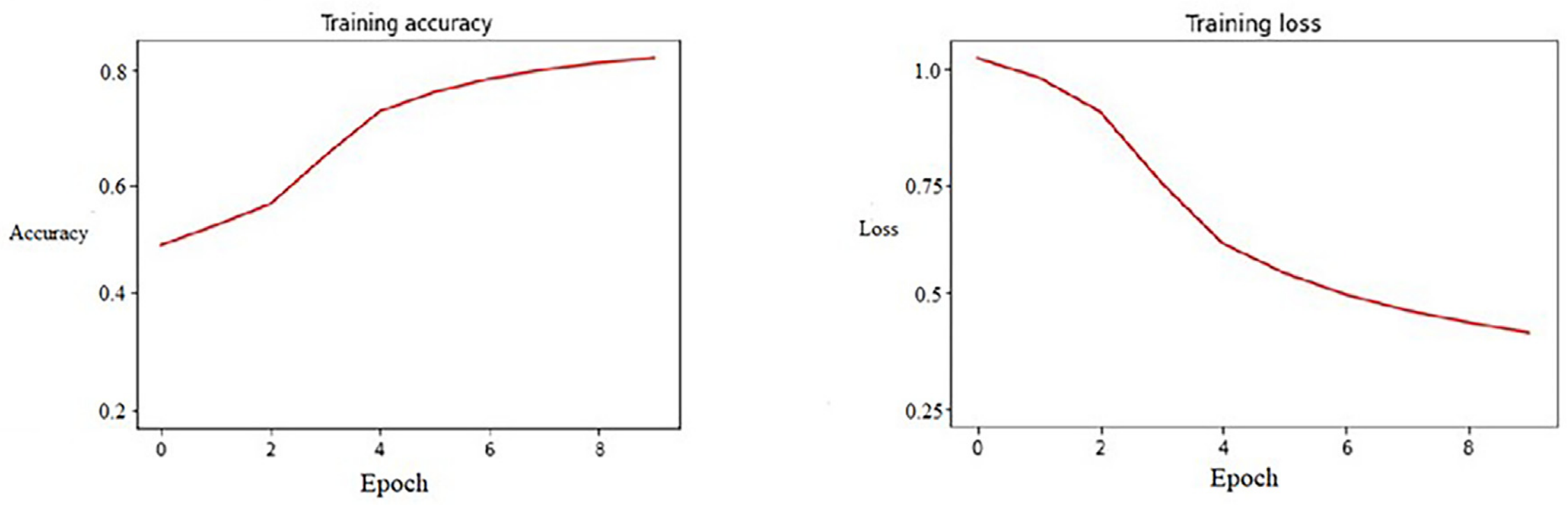

**Fig. 7.** Graphs depicting the training accuracy and loss with each epoch.

### 5.13. Model setup

Posts of Reddit users are attained using PRAW through the Reddit API and passed through the model to detect whether it is suicidal or not (Tanner, 2019). A user id of a particular Reddit user was given as input and the 10 most recent posts of the user was returned, which was then passed through the model to generate labels for the individual posts. In order to provide a label to the user we converted the labels into numeric values and then applied exponential weighted moving average on the label values. This would then generate a particular label, or numeric value which corresponded to a particular label, for a user.

### 5.14. Reddit API

The Reddit API allows us to access information from Reddit in json format and PRAW is a Python wrapper for the Reddit API, which enables us to scrape data from subreddits. For using the Reddit API to get posts from a user, we first need to register or create an account on Reddit and then create an application on Reddit by filling in the basic details like name, app type (script) and redirect URL. On submitting the app details, we get a public key and a secret key which is required to use the Reddit API. Then we create a Reddit instance and provide it with a client id (provided under personal use script), client secret (the secret value) and a user agent (app name provided). Now we can make requests for posts belonging to the user id we specify. The posts of a particular user are then passed through our proposed model to predict the label for each post, and then, the exponential weighted moving average is employed to give a label for the user.

### 5.15. Exponential weighted moving average

Exponentially weighted moving average (EWMA) can be defined as a quantitative or statistical measure where we represent or model the given time series data in such a way that the older measures or values are given lesser weights. The weight for each value decreases exponentially as the older the data gets. Here we provide a value which depicts how important our current or latest observation is when calculating the EWMA. In our model we divide the weighted value by the decaying adjustment factor in beginning periods in order to account for the imbalance in relative weightings. Let the data be [x1, x2, . . ., xn] then the EWMA can be represented as in equation (9).

$$y_t = \frac{(x_t) + (1-\alpha)x_{t-1} + (1-\alpha)^2 x_{t-2} + \cdots\cdots + (1-\alpha)^t x_0}{1 + (1-\alpha) + (1-\alpha)^2 + \cdots + (1-\alpha)^t} \quad (9)$$

where $\alpha$ is the weight. Based on the sensitivity analysis conducted as part of our experiments we defined $\alpha$ = 0.769 in this research.

### 5.16. Evaluation metrics

The performance of our model is estimated mainly using accuracy metric. In this case, accuracy is the fraction of predictions of our model that was correctly classified. In addition to accuracy, we also find the recall, F1 Score and Precision. Precision can be defined as the proportion of the positive posts that were actually correct. Recall is the proportion of the actual positives that were identified correctly. The F1 Score combines recall and precision using harmonic mean.

## 6. Experimental results

We started by exploring the results of data analysis from our complete list of Reddit posts. The main objective of this work is the detection of suicidal tendencies from the selected data. We examined the Severe Risk labeled posts associated with suicidal tendency and created a list of the most common n-grams and then compared this with those obtained from Low-Risk labeled posts. We used features of our proposed model to detect the indications of suicidal thoughts. We trained our data for 10 epochs and our model got an accuracy of 90.3%. The graphical representation of the accuracy and loss at each epoch is shown in Fig. 7.

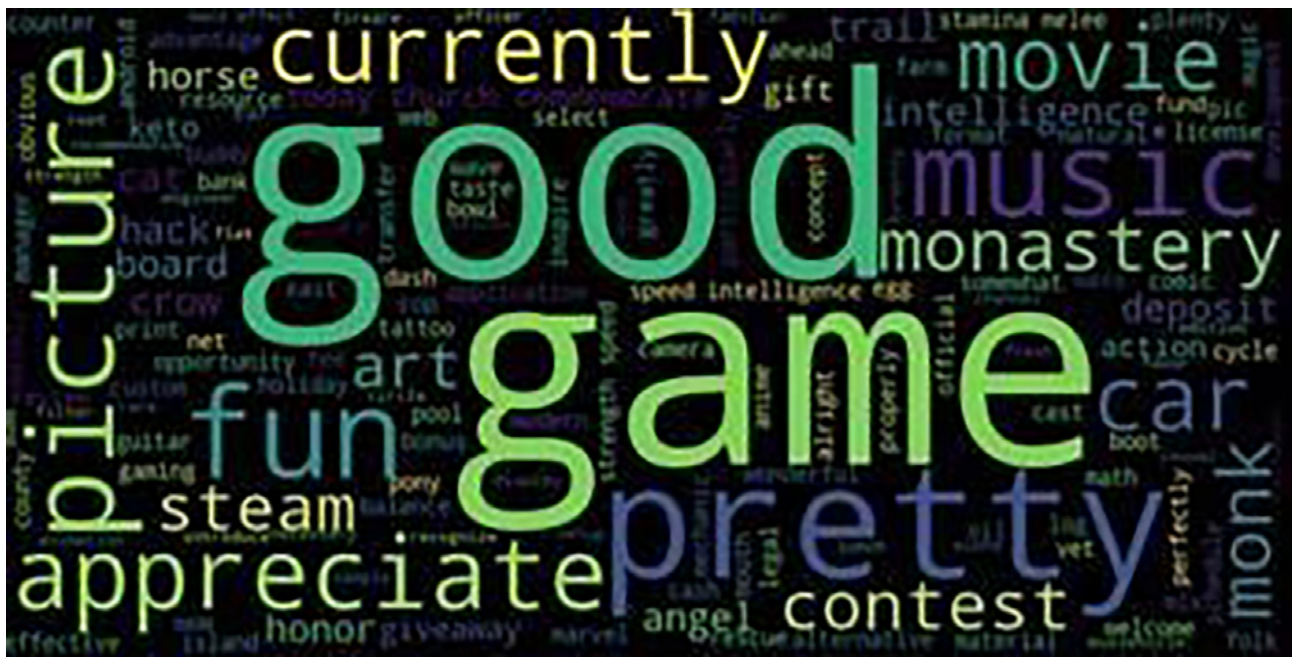

**Fig. 8.** No risk.

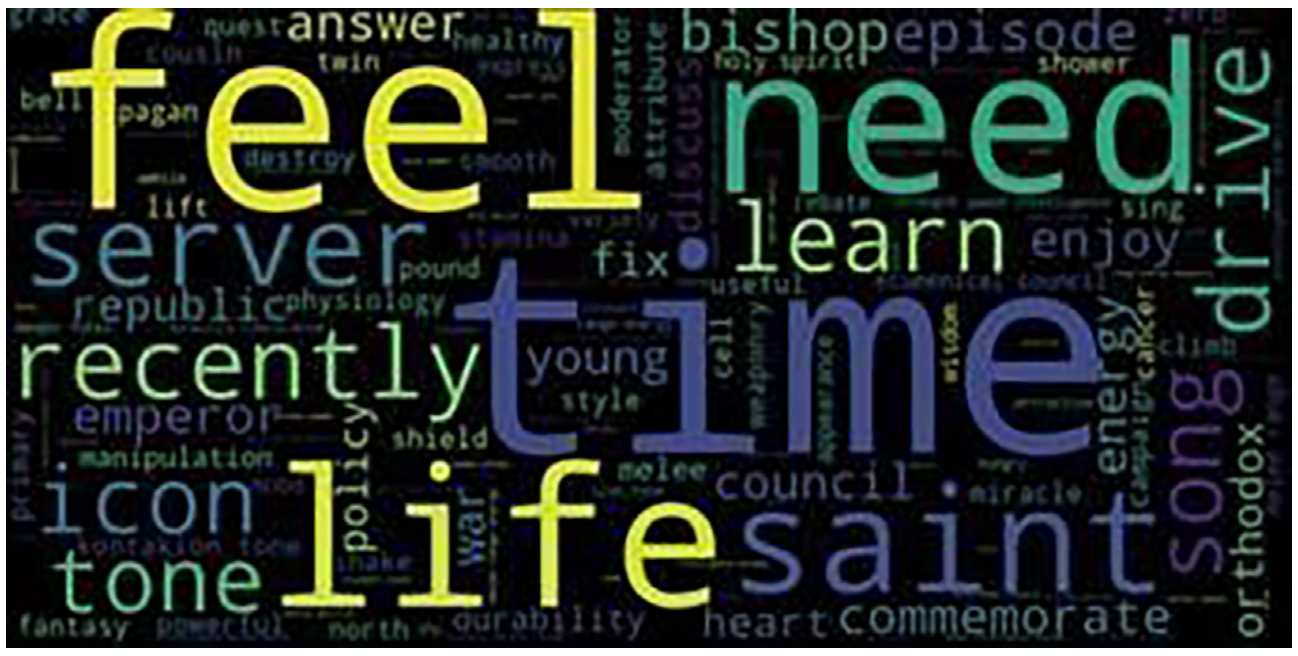

**Fig. 9.** Low risk.

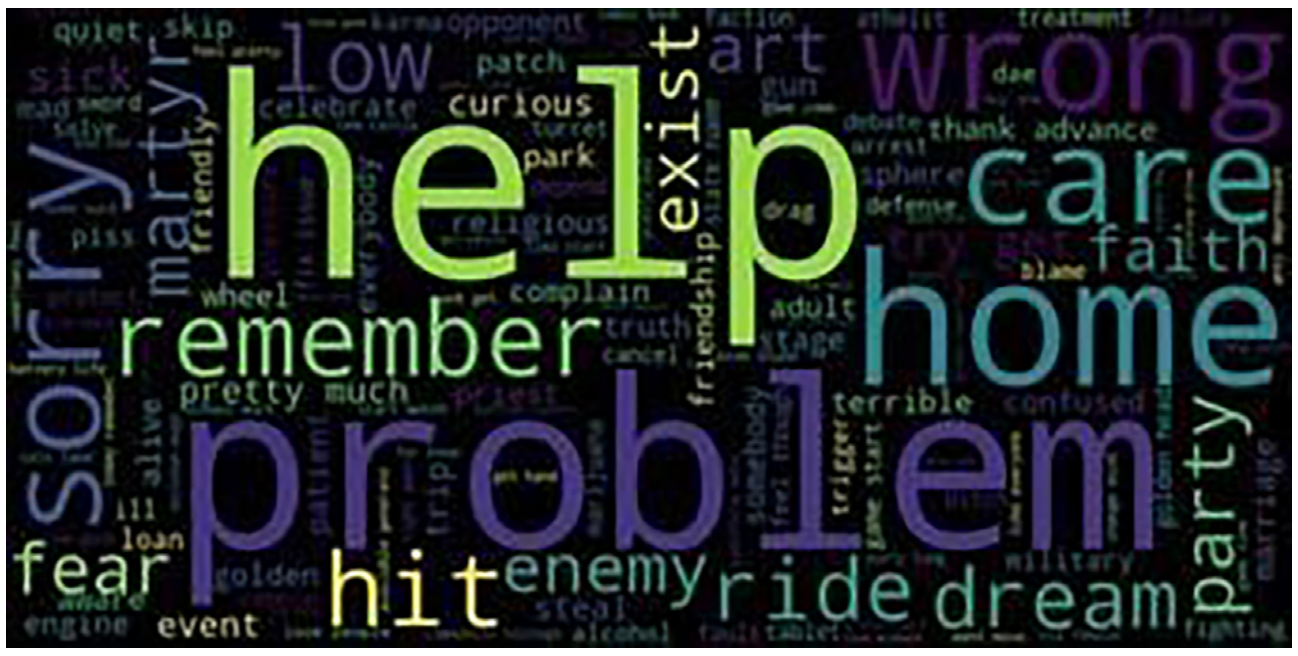

**Fig. 10.** Moderate risk.

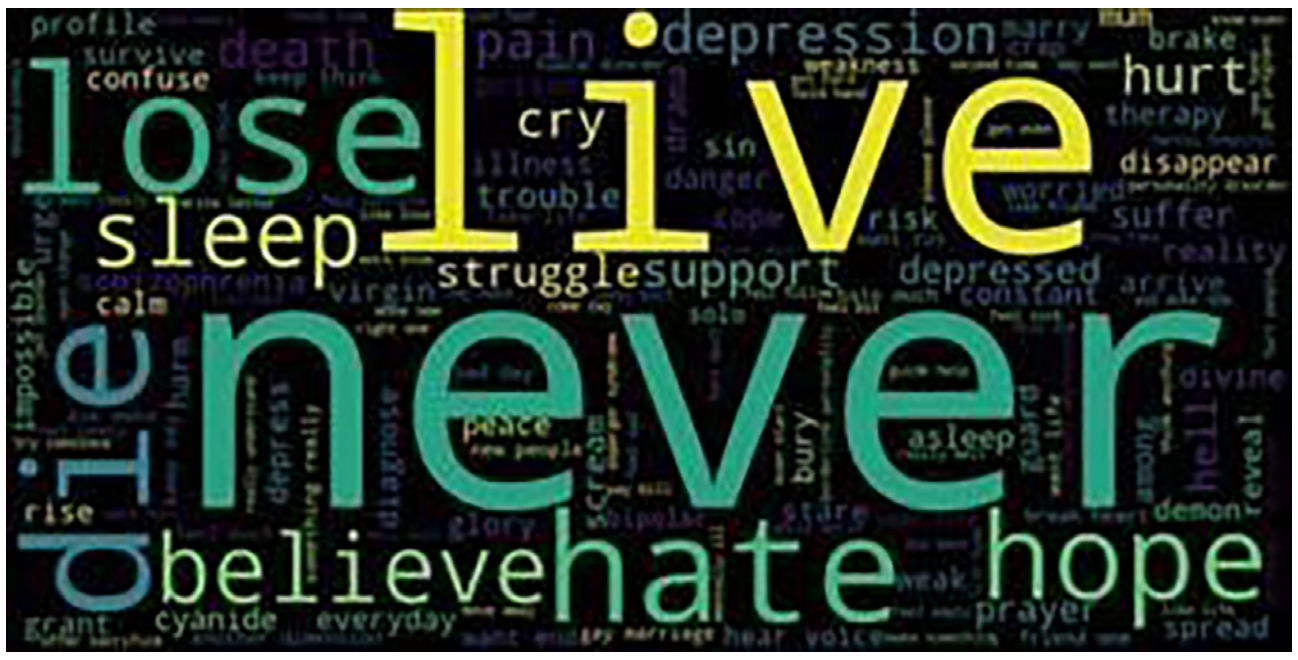

**Fig. 11.** Severe risk.

## 6.1. Data analysis result

We inspected the whole input dataset in order to identify posts with suicidal ideations, and to compare dissimilarities in the word lists. The frequencies of all the unigrams, bigrams, and trigrams in the four different classes of our model were computed. The top 300 unigrams, bigrams and trigrams belonging to the specific groups were selected to study their relationship with suicidal thoughts. Visual support of the word cloud was used for our understanding of frequently occurring words in four different classes which are shown in Figs. 8–11.

From the Reddit SuicideWatch forum, we examined posts of various users and recognized evidence indicating frustration and hopelessness like ("never", "impossible", "I'm tired"), the feeling of guilt ("sorry"), fear ("scream"), disappointment ("fail follow"), solitude ("feel lonely"). We also observed that the attention or the focus of the social media users was directed towards themselves. Next, we identify the person's inclination for the preoccupation with their sentiment ("really understand", "feel much"), backed by the expressions of contradiction ("no one", "do not", "would never"). From the Severe Risk labeled posts, we observed the usage of a large number of question marks ("Troubled but don't know anything to do?", "Why are people afraid of dying?").

Users' depiction of suicidal tendencies was another interesting observation which was mainly conveyed by the terms having death implications ("death", "die", "disappear", "want cut", "die want", "want end", "want hurt"). Furthermore, a feeling of haste, as well as an indication of desperation, is also seen ("wish help", "feel empty", "move now"). The n-grams observed in the No Risk labeled posts consist of terms expressing happiness, positive mindset, and emotions ("need joke", "fun", "good game", "appreciate") which is entirely opposite to Severe Risk posts. The users tend to maintain their positive spirits ("make better") and usually state social relations actions ("friend", "high school").

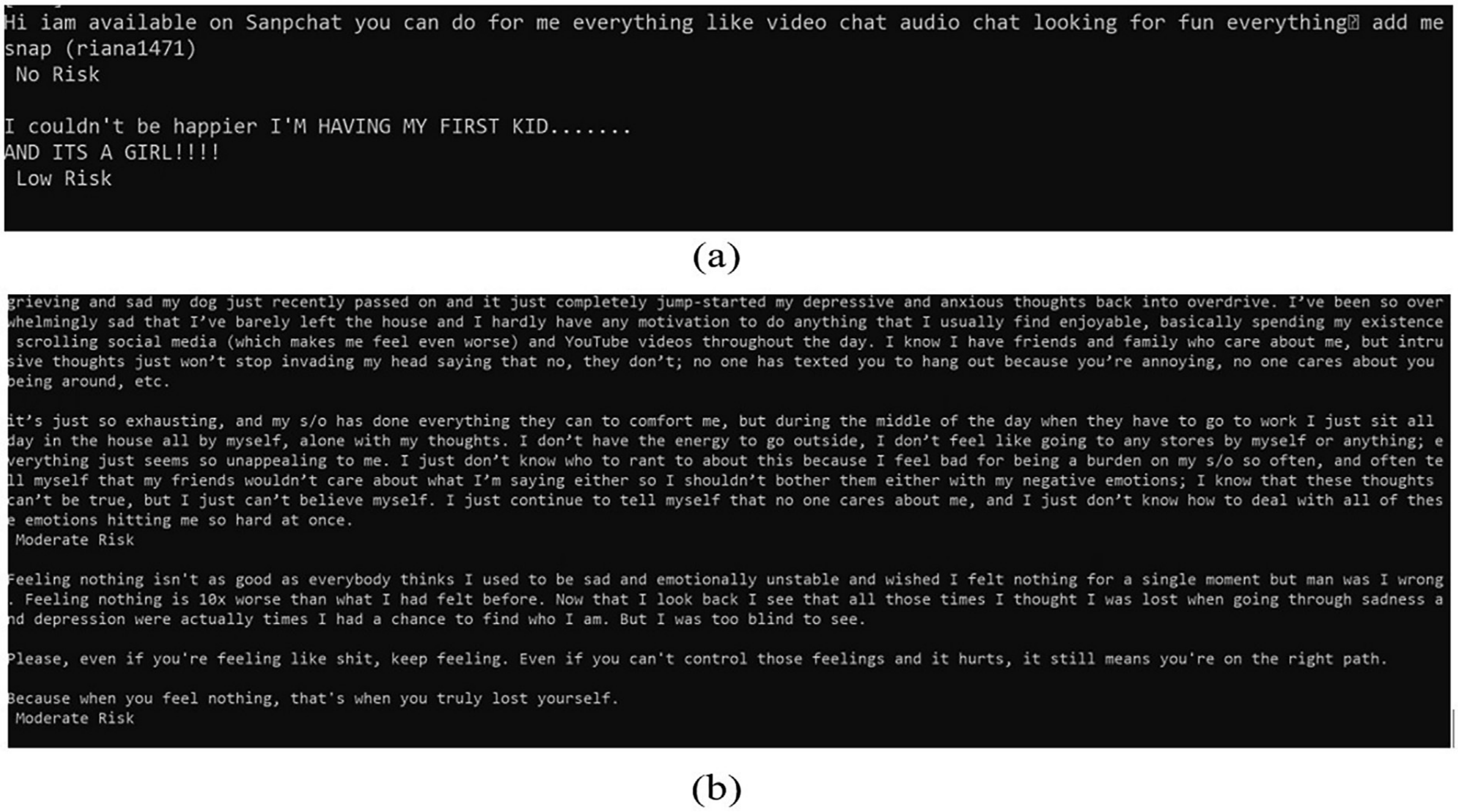

**Fig. 12.** Result depicting predicted value of posts.

**Table 4**
Performance results of the classification model.

| Models | Accuracy | Precision | Recall | F1-Score |
|---|---|---|---|---|
| SVM | 82.5 | 81.1 | 84.5 | 82.8 |
| CNN | 86.6 | 87.8 | 89.8 | 88.8 |
| LSTM | 87.7 | 90.8 | 86.5 | 88.6 |
| LSTM-CNN | 88.2 | 88.3 | 84.4 | 86.3 |
| Proposed Model | 90.3 | 91.6 | 93.7 | 92.6 |

### *6.2. Classification result analysis*

We employ the word embedding layer using word2vec as the input layer for our experiment. The output of this layer is given as input to the different baseline models to perform classification. The model is trained for 10 epochs the value of which was selected after going through various related models. Early stopping is another technique which could be used to stop training upon satisfying a condition. Some predictions made by the proposed model are shown in Fig. 12. The figure shows posts or submissions taken from Reddit with the corresponding label below each post. Fig. 12 (a) shows a post labeled as No Risk which depicts a user wanting to start a conversation for fun and a post wrongly labeled as Low Risk of a happy person who just became a parent which is a No Risk post. Fig. 12(b) has two posts labeled Moderate Risk, which shows the emotions of two individuals.

The various evaluation metric values are obtained for the different baseline models and are shown in the Table 4 below. Upon comparing our proposed model with the other baseline models, we observe that the best classification in our experiment is achieved with LSTM-attention-CNN combined model. With the optimized parameters, it significantly outperformed the other models and demonstrated an accuracy of 90.3%.

We also understand that accuracy cannot be a proper metric when the dataset is imbalanced. So, we also evaluated the models using more accurate measures like precision, recall and F1-score. The proposed model demonstrated a precision of 91.6%, recall of 93.7% and F1-score of 92.6%. These results demonstrate that the proposed combined model performs better than the single LSTM and CNN classifiers and the combined LSTM-CNN model.

Macro-averaging is used to calculate precision, recall and accuracy of the proposed model. We observe that the precision and recall values of the proposed model are found to be greater than the accuracy. This can be due to the high recall values of severe risk and no risk classes and comparatively lower values for low risk and moderate risk classes and also the high precision values of moderate risk and severe risk classes. The good precision and recall of these classes contribute to maintaining a good overall precision and recall but a lower accuracy.

## 7. Conclusion and future work

In our paper, we have designed and tested an approach to check the mental health of different Reddit users. We have developed this model intending to reduce the rate of suicide in the society by detecting suicidal intentions from social media interactions of individuals. This can help to provide proper medical assistance and/or help to those in need of the same. Based on the results obtained from our model, we can also provide an awareness class on how to deal with mental issues or stress, anxiety, etc. Nowadays, people are more sensitive to handle simple situations, to deal with hardships, etc. which brings out great losses to families, friends, and those around them. Thus, we felt the need to develop a model which will reduce the rate of suicides by detecting suicidal related posts in social media.

Using the LSTM-attention-CNN combined model we were able to analyze submissions by giving importance to the regions in the data that depicted whether the submission was suicidal or not. With an accuracy of 90.3% and F1-score of 92.6%, the performance of our proposed model was also observed to be greater than the other baseline model.

Our future work aims at experimenting with different attention mechanisms, especially the Hierarchical Attention Networks (HAN) to see how structuring of documents could affect the performance of our proposed classification model.

## Declaration of Competing Interest

The authors declare that they have no known competing financial interests or personal relationships that could have appeared to influence the work reported in this paper.

## Acknowledgments

We acknowledge support from the Department of Computer Science and Engineering, Mar Baselios College of Engineering and Technology for all the support being extended to carry out this research work. We also acknowledge the help extended by Professor Philip Resnik for providing the dataset, The University of Maryland Reddit Suicidality Dataset, Version 2.